\documentclass[11pt]{article}
\usepackage{sao1}
\usepackage{graphicx}

\begin{document}

\title{Spectropolarimetry with the DAO 1.8-m telescope
\thanks{Based on observations acquired at the Dominion Astrophysical
Observatory, Herzberg Institute of Astrophysics, National Research Council
of Canada.}
}
\author{Dmitry Monin \and   David Bohlender\inst{}}
\institute{Herzberg Institute of Astrophysics, National Research Council of
Canada\\
5071 West Saanich Road, Victoria BC, Canada V9E 2E7}

\maketitle 

\begin{abstract}
The fast-switching DAO spectropolarimeter mounted
on the 1.8-m Plaskett telescope started operation in 2007.
Almost 14,000 medium-resolution ($R\approx 15,000$) polarimetric spectra
of 65 O - F type stars have been obtained since then
in the course of three ongoing projects:
the DAO Magnetic Field Survey, supporting observations for
the CFHT MiMeS survey, and an
investigation of the systematic differences between
the observed longitudinal field measured with the
H\,$\beta$ line and metallic lines.
The projects are briefly described here.
The current status as well as some results are presented.
\end{abstract}

\section{The DAO Spectropolarimeter}
The Dominion Astrophysical Observatory (DAO) is located on Vancouver Island,
just north of Victoria, British Columbia. The Observatory is operated by
the National Research Council of Canada's Herzberg Institute of Astrophysics and
has two telescopes located on its grounds:
the 1.2-m and the 1.8-m Plaskett telescope. The 1.2-m telescope is
used exclusively for high-resolution coud\'{e} spectroscopy, often in robotic mode.
The 1.8-m Plaskett telescope is used for direct imaging, spectroscopy,
and spectropolarimetry.

A grating spectrograph that offers a wide variety of spectral resolutions
ranging from $R=200$ to 15,000 is mounted at the Cassegrain focus
of the 1.8-m Plaskett telescope.
Three years ago an inexpensive polarimetric module for this
spectrograph was designed and built by observatory staff (Monin et al.~2010).
The module is installed behind the spectrograph slit.
The polarimetric module and the Cassegrain spectrograph together form what is
known as the DAO spectropolarimeter.
The instrument is optimized to measure circular polarization
from stellar sources in a 260\,\AA\ wide spectral window centered on H\,$\beta$
with a resolution of about $R=$15,000.

\begin{figure*}
\centerline{\includegraphics[angle=270,width=120mm]{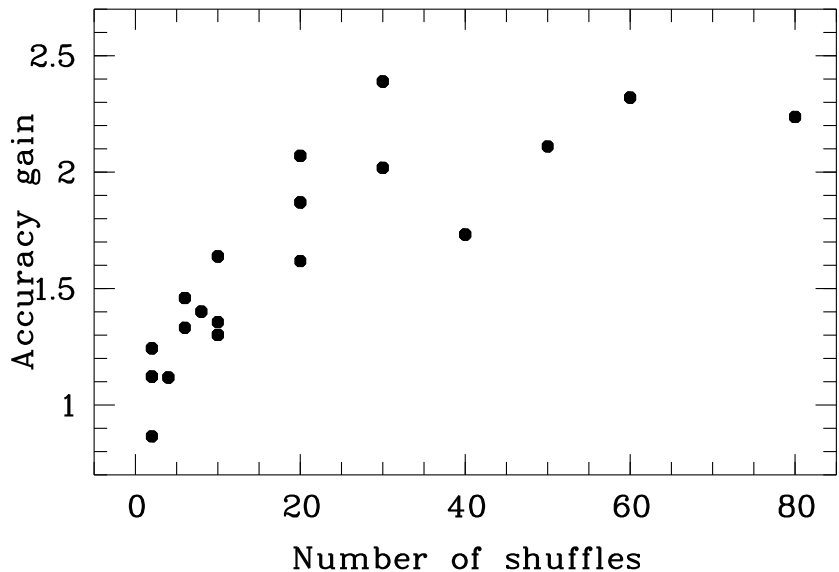}}
\caption{The DAO spectropolarimeter uses fast modulation
in order to fight instrumental effects. The relative increase in positional
accuracy versus the number of switch cycles is shown in this figure.
With 20 or more switch cycles the accuracy improves by a factor
of two or more.}
\label{accuracy}
\end{figure*}

The module incorporates a two-beam design. An achromatic quarter-wave
plate converts circularly polarized light into linear. A calcite beam displacer
then separates the orthogonal linear polarization components. The two beams travel parallel
to each other inside the spectrograph. As a result, two spectra in opposite
polarizations are recorded on a CCD.
Since both spectra are taken under exactly the same conditions,
the polarization signal is independent of sky transparency, seeing or slit
losses but not the instrumental systematics.
In our spectropolarimeter we use fast modulation to fight instrumental
systematics. A fast-switching ferro-electric liquid crystal (FLC) half-wave
plate is installed between the quarter-wave plate and the beamsplitter.
The FLC half-wave plate interchanges the two beams so that
the beams travel the same path inside of the instrument.
The waveplate is synchronized with shuffling of the charge back and forth
on the CCD (Chountonov et al.~2000).
The DAO spectropolarimeter can be switched up to 100 times per second.
In Fig.\,\ref{accuracy} the gain in positional accuracy is plotted against
the number of switch cycles.
By executing 20 or more switch cycles (we typically switch 60 times during each exposure) the accuracy can be improved
by a factor of two or more.
Another advantage of this modulation approach is that no flat-fielding
is necessary since the same CCD pixels are used for both polarizations.

\begin{figure*}
\centerline{
\includegraphics[angle=270,width=80mm]{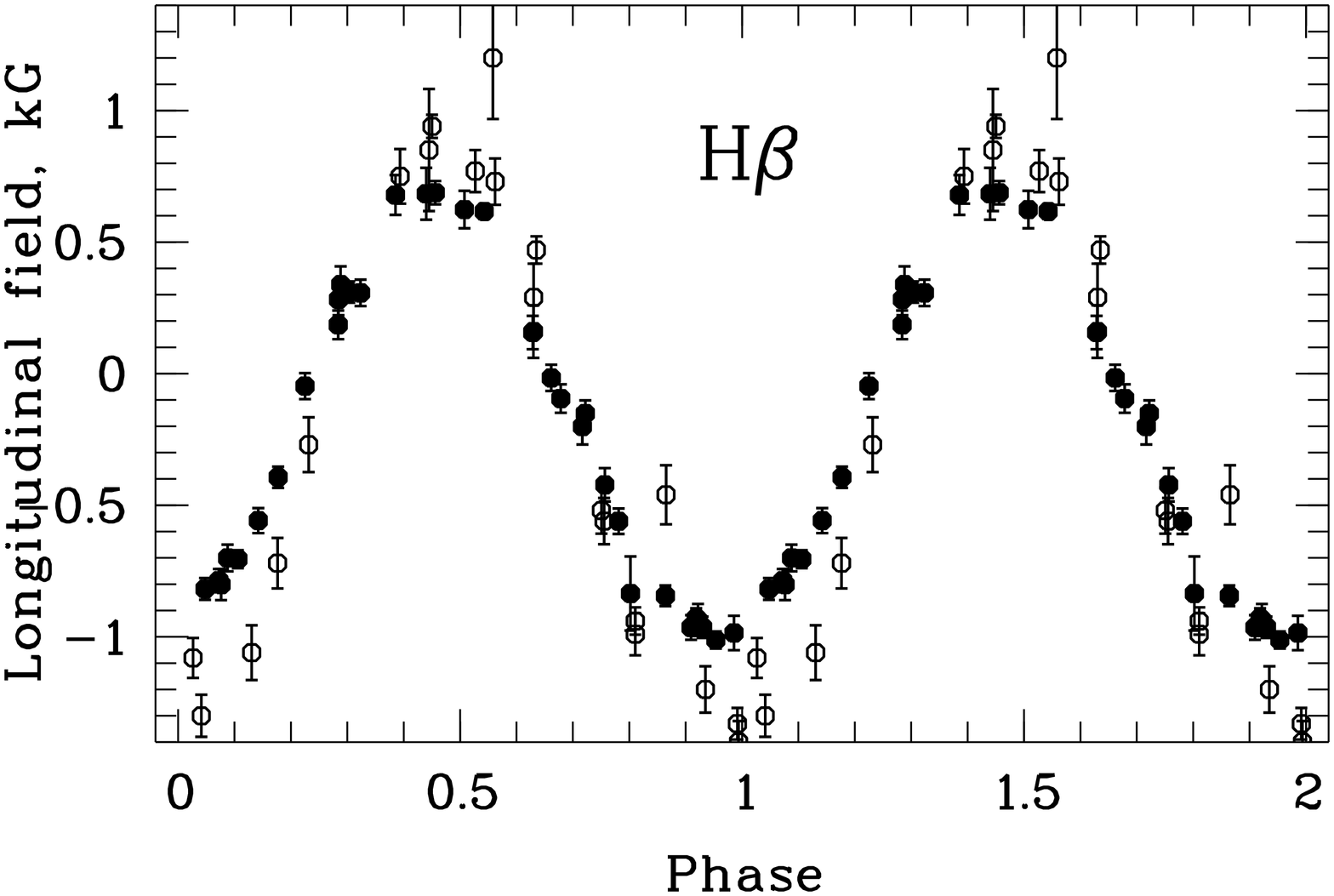}
\includegraphics[angle=270,width=80mm]{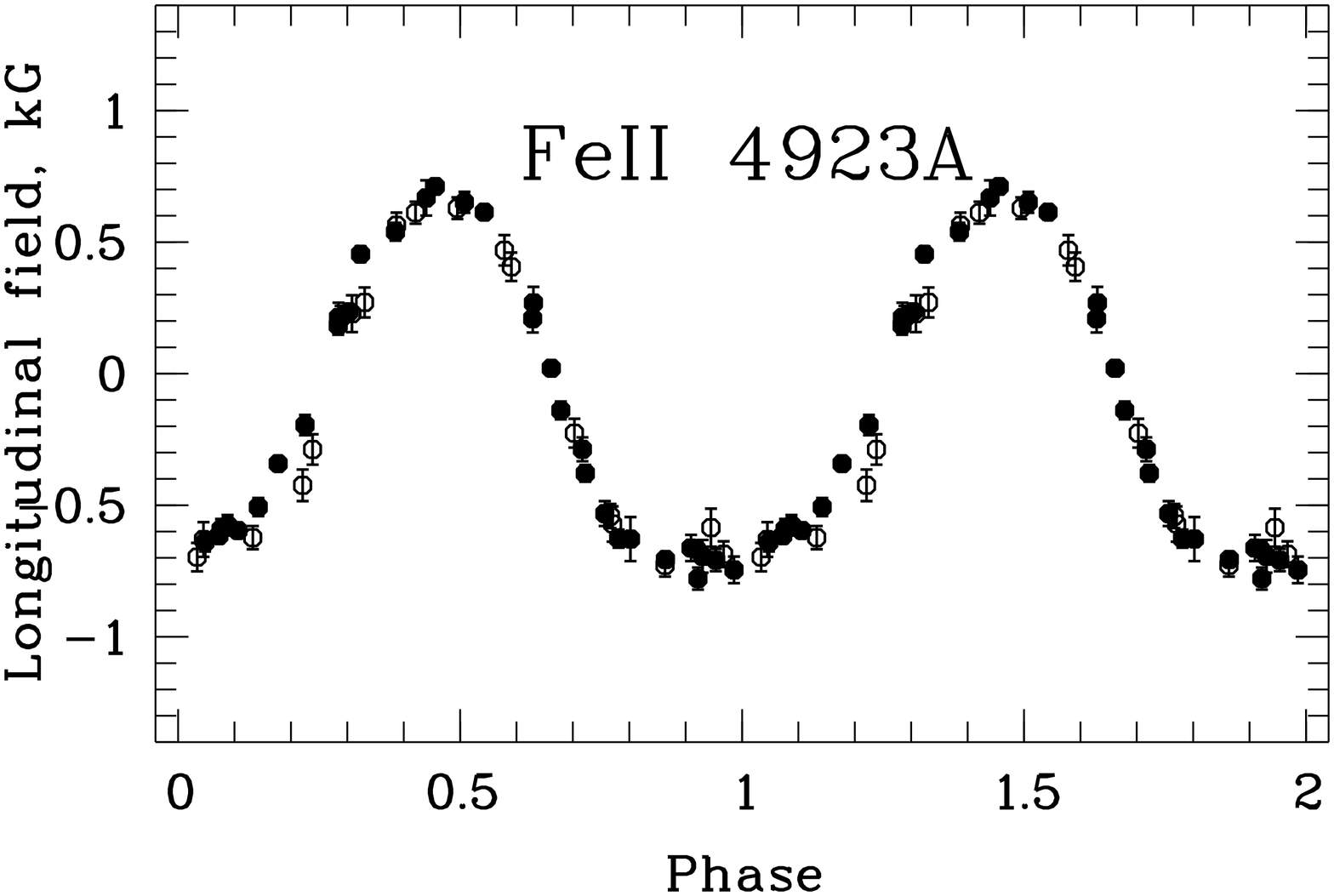}
}
\caption{Longitudinal field measurements of the well-known
magnetic star $\alpha^2$\,CVn.
Left graph: our H\,$\beta$ line measurements (filled circles)
and the Balmer line photoelectric measurements
by Borra and Landstreet~(1977) (open circles).
Right graph: our Fe\,II\,$\lambda$4923 line measurements
and the metallic line LSD measurements by Wade et al.~(2000).}
\label{alp2cvn}
\end{figure*}

The DAO spectropolarimeter was mainly designed to measure longitudinal
magnetic fields in stars through use of the Zeeman effect. In a magnetic field
spectral lines split into two or more components and the individual components
have different polarizations. The DAO spectropolarimeter can separate
components in different polarizations so that the small magnetic shift
can be measured.
Magnetic shifts are usually quite small, less than a single CCD pixel.
Great care is needed in order to measure such small line displacements.
We apply Fourier cross-correlation technique to measure the shift.
The shift is usually measured in the line core since the polarization signal
is strongest there, although other parts of the line profile can be used as
well.
Magnetic field measurements in the wings of H\,$\beta$ agree well
with line core measurements if the appropriate correction for Stark effects
is applied to the line wing measurements (Mathys et al.~2000).
The accuracy of line core measurements is usually higher
however. The DAO spectropolarimeter enables us to measure the magnetic shift
in both the hydrogen H\,$\beta$ as well as metallic lines. This is a big advantage of
our spectropolarimeter over some other instruments.
Fig.\,\ref{alp2cvn} shows how our measurements compare with previously
published data for the bright well-known magnetic
star $\alpha^2$\,CVn.
Our H\,$\beta$ measurements agree with
the Balmer line photoelectric measurements obtained
by Borra and Landstreet~(1977) while our measurements of the Fe\,II $\lambda$4923  line
agree perfectly with the metallic line measurements obtained by Wade et al.~(2000).

\section{Spectropolarimetry with the DAO 1.8-m Telescope}

The DAO spectropolarimeter was commissioned in October 2007. Since then
it has been used extensively for several different projects.
The spectropolarimeter is typically scheduled for two to three weeks of observing time each quarter.
In three years of operation 519 magnetic field measurements have been
obtained for 65 O - F type stars as faint as 9th magnitude.
The instrument archive now contains 13,728 individual polarization spectra.
Both slowly and rapidly rotating stars have been successfully observed,
with the fastest object rotating at about 400\,km\,s$^{-1}$.
An accuracy of about 20 to 100\,G in H\,$\beta$
can be achieved with the DAO spectropolarimeter; this depends mostly
on brightness of the object and its $v \sin{i}$.

The DAO spectropolarimeter is used in three major ongoing projects:

\begin{itemize}
\item MiMeS support program
\item DAO Magnetic Field Survey
\item Hydrogen versus metallic line magnetic field measurements
\end{itemize}

The first project is support observing for the CFHT Large Program Magnetism
in Massive Stars (MiMeS).
The MiMeS program is aimed at significantly improving our knowledge of
how magnetic fields in massive stars influence massive star evolution
and also how the fields modify mass loss in these stars
(see Gregg Wade's article in these Proceedings).
Part of this program consists of a large survey of stars more massive than
8 solar masses. At the DAO we have been doing follow up spectropolarimetric
observations of some objects from the MiMeS survey.

The DAO Magnetic Field Survey is an extension to the MiMeS survey
to less massive upper main sequence stars. Many of these stars,
especially fainter ones, have not been examined for the presence
of a magnetic field. Most well-known magnetic B-type stars with magnetospheres
are non-thermal radio sources. In the early stages of the DAO Magnetic Field
Survey we therefore were concentrating our efforts on observing peculiar
B type stars known to be radio sources. As a result we have discovered four new
magnetic stars: HD\,135679, HD\,164429, HD\,176582, and HD\,189775.
HD\,176582 (B5IV, V=6.4, $v\sin{i}=100$\,km\,s$^{-1}$) was the first magnetic star
discovered at the DAO (Bohlender and Monin~2010).
Since then we have broadened our target list considerably to now include non-radio
sources. The list also includes many cooler Ap type stars.

\begin{figure*}
\includegraphics[angle=270,width=80mm]{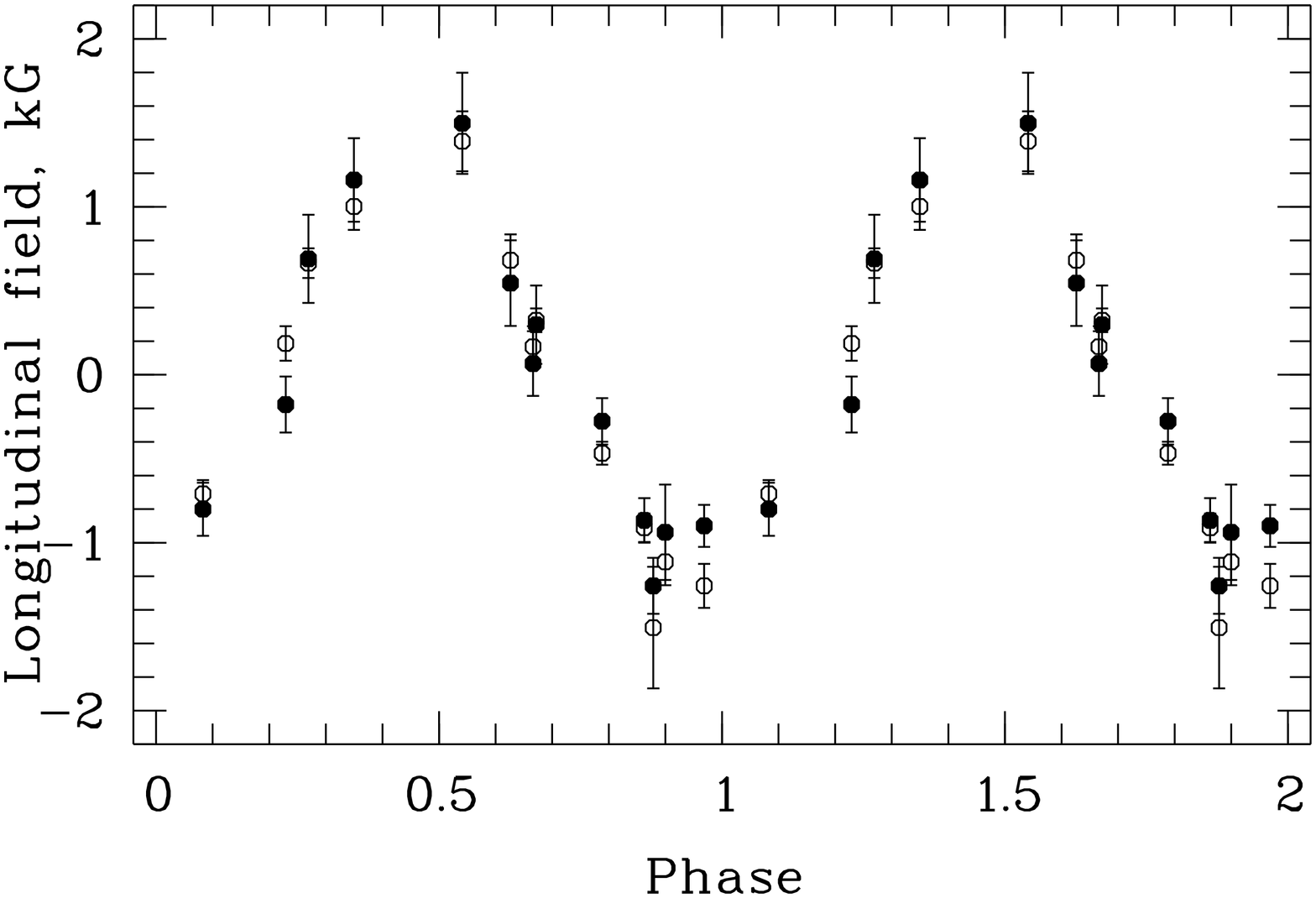}
\includegraphics[angle=270,width=80mm]{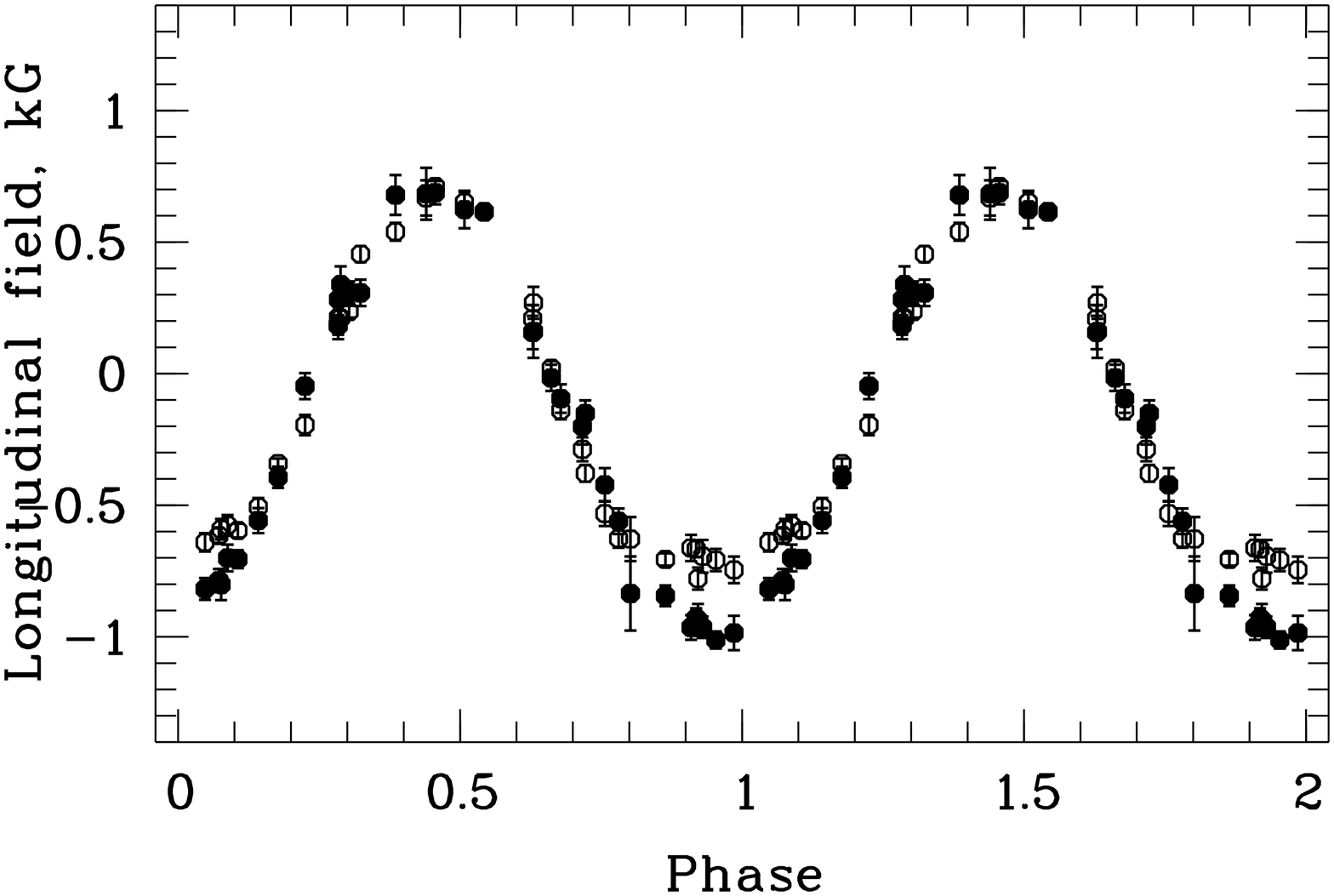}
\includegraphics[angle=270,width=80mm]{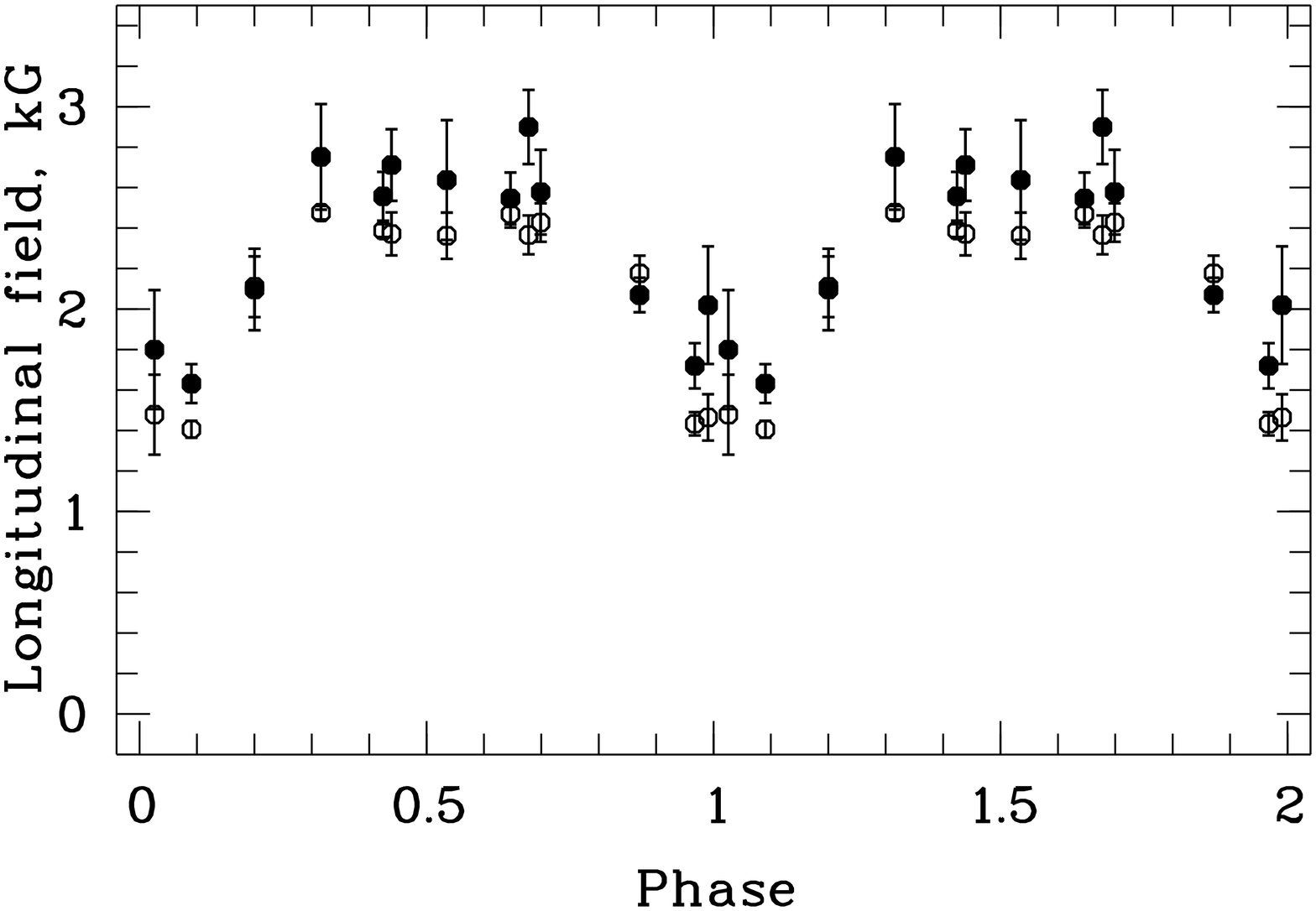}
\includegraphics[angle=270,width=80mm]{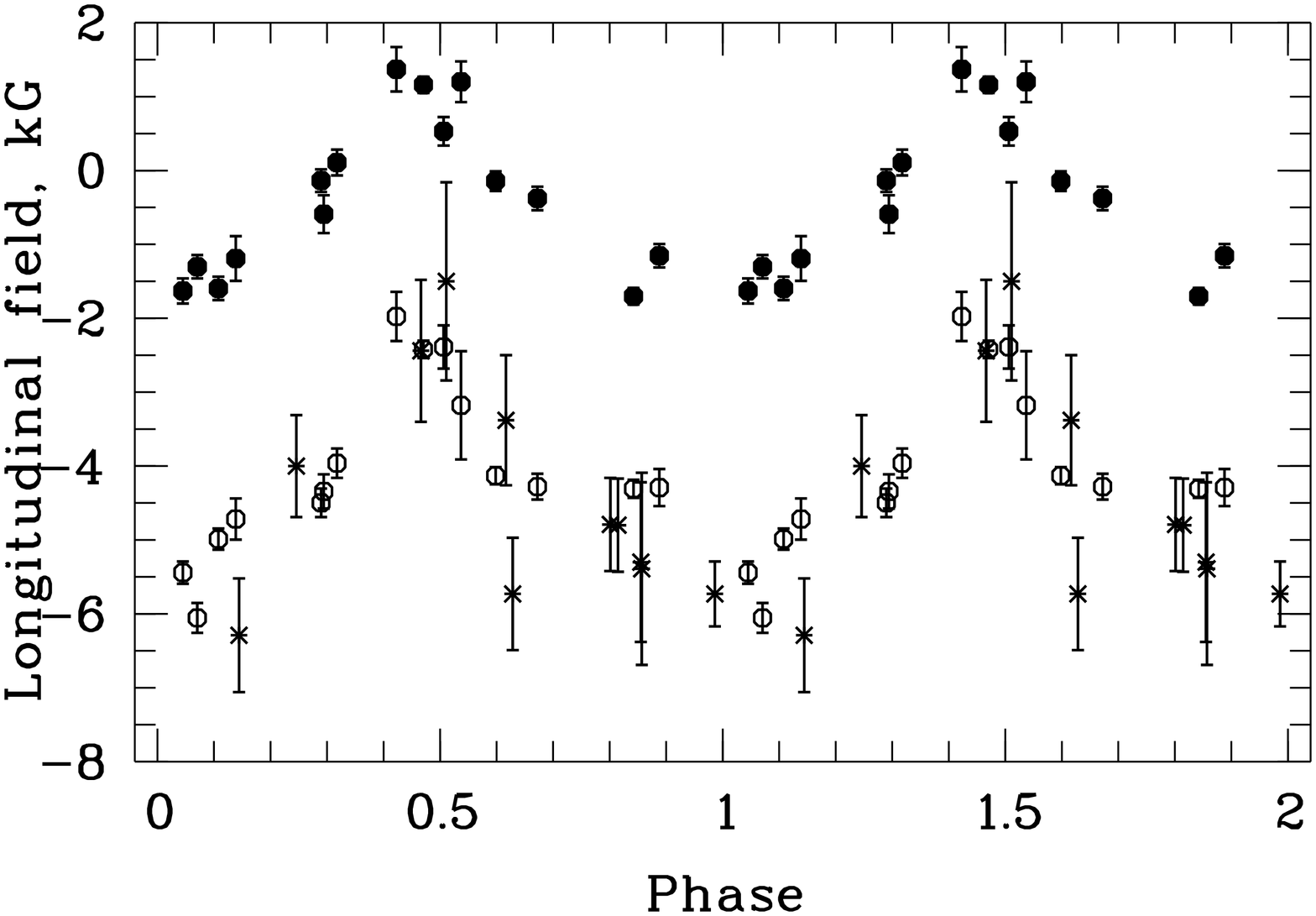}
\caption{Hydrogen and metallic line longitudinal magnetic field curves
obtained with the DAO spectropolarimeter for four known magnetic stars.
Filled circles: hydrogen H\,$\beta$;
open circles: Fe\,II~$\lambda$4923;
asterisk: metallic line measurements by Elkin~1994.}
\label{hbeta_metals}
\end{figure*}

Our third program aims to investigate the long-standing puzzle
of differences observed between magnetic field measurements obtained
using Balmer lines versus those acquired with metallic lines.
These differences are commonly attributed to inhomogeneous distribution
of metals over the stellar surface.  However, differences in
observational techniques and/or interpretation of data
may also play a role since, due to a big difference in line widths, hydrogen lines
and metallic lines are usually observed with different instruments.
The DAO spectropolarimeter is capable of measuring both hydrogen
and metallic lines simultaneously and
we have therefore observed several stars with well-established magnetic fields
and well-known periods throughout their rotation cycles.
For some stars we see no difference between H\,$\beta$ and
Fe\,II $\lambda$4923 line measurements
(see upper left panel of Fig.\,\ref{hbeta_metals}).
For other stars there is a noticeable difference only at certain phases
(Fig.\,\ref{hbeta_metals}, upper right), while
some stars show an offset between the hydrogen line and
iron line curves
(Fig.\,\ref{hbeta_metals}, bottom left).
One star, HD\,217833, shows a huge offset of about 4\,kG
(Fig.\,\ref{hbeta_metals}, bottom right).
Our iron line measurements for this star are in a good
agreement with previously published metallic line measurements (Elkin~1994).
A simple model of these metallic line measurements suggests that
the magnetic dipole is almost aligned with the rotation axis
and that the polar magnetic field is one of the strongest
among upper main sequence stars.
Our hydrogen line measurements, however, suggest a totally different picture.
The dipole is tilted almost 90$^{\circ}$ to the rotation axis and the polar field strength is four
 times less than what is derived
from the metallic lines. 
The most likely reason for the large difference in the two curves
is that iron is concentrated
in a small spot over the negative magnetic pole.
A paper outlining our findings is now in preparation.
What is obvious is that hydrogen line measurements are needed for
proper characterization of a star's magnetic field geometry and that
the DAO polarimeter can provide this kind of information.

\section{Summary}

The DAO spectropolarimeter mounted on the 1.8-m Plaskett telescope is capable
of measuring longitudinal magnetic fields in both hydrogen H\,$\beta$
and metallic lines.  Both slowly and rapidly rotating stars can be studied with the instrument.
The data gathered will allow us to look into the systematic differences
between Balmer and metallic line magnetic field measurements.

Nearly 14,000 spectra of 65 O - F stars have been obtained since
the instrument was commissioned in 2007.   Four new magnetic stars have been
discovered.   Periods have been obtained or revised for a number of the observed stars.

A MiMeS support program as well as a DAO Magnetic Field Survey of upper main sequence peculiar
stars are well underway.  Coordination of observations with astronomers conducting similar surveys at SAO
or other locations would be highly beneficial.

\end{document}